\documentclass[a4paper,twocolumn]{article}

\usepackage[T1]{fontenc} 
\usepackage{float}       
\usepackage{helvet}      
\usepackage{mathptmx}    
\usepackage{rsc}         
\usepackage{setspace}    
\usepackage{graphicx}    

\newfloat{chart}{htbp}{loc}
\newfloat{graph}{htbp}{loh}
\newfloat{scheme}{htbp}{los}

\newcommand{\dir}{Figs}

\newcommand{\citeonline}[1]{\citep{#1}}
\newcommand{\cites}[1]{$^{\mbox{\small\citealp{#1}}}$}

\author{%
  Beate West $^1$
  and
  Friederike Schmid $^{1,2}$ \\
  ${^1}$ Physics Faculty, University of Bielefeld, D-33615
  Bielefeld \\
  ${^2}$ Institute of Physics, University of Mainz, D-55099 Mainz,
  Germany
}

\title{Fluctuations and Elastic Properties of Lipid Membranes
in the Gel $L_{\beta'}$ State: A Coarse-Grained Monte Carlo Study}

\begin{document}

\maketitle

\begin{abstract}

We study the stress distribution profiles and the height and thickness 
fluctuations of lipid membranes in the gel $L_{\beta'}$ state by Monte 
Carlo simulations of a generic coarse-grained model for lipid membranes, 
which reproduces many known properties of dipalmitoylphosphatidyncholine 
(DPPC) bilayers. The results are related to the corresponding properties
of fluid membranes, and to theoretical predictions for crystalline 
and hexatic membranes. One striking observation is that the spontaneous 
curvature of the monolayers changes sign from the fluid to the gel phase. 
In the gel-phase, the long-wavelength height fluctuations are
suppressed, and the fluctuation spectrum is highly anisotropic.
In the direction of tilt, it carries the signature of soft modes 
that are compatible with the wavelength of the ripple phase
$P_{\beta'}$, which emerges in the transition region between
the fluid and the gel state. In the direction perpendicular to
the tilt, the thickness fluctuations are almost entirely 
suppressed, and the height fluctuations seem to be dominated 
by an interfacial energy, {\em i.e.}, by out-of-layer fluctuations, 
up to length scales of tens of nanometers.
\end{abstract}

\section{Introduction}

\label{sec:introduction}

Lipid molecules in water self-assemble into a variety of nanostructures. 
Among the most prominent are bilayers, where the lipids are arranged 
such that their hydrophobic hydrocarbon tails are shielded from 
the water by the hydrophilic head groups. Lipid bilayers form the matrix
for biological membranes, and they are ubiquitous in all 
living matter \cites{Gennis89}. 

Model systems of pure (one-component) lipid membranes in water have 
been studied extensively for many decades
\cites{BEM91,KC94,KC98}. Depending on
the temperature and on the structure of the lipids, one finds a variety
of phases: A high-temperature ``fluid'' phase ($L_{\alpha}$) 
where the lipid tails are highly disordered, a low-temperature ``gel''
phase with either straight ($L_{\beta}$) or collectively tilted 
($L_{\beta'}$) tails, and in the latter case an intermediate modulated 
``ripple'' phase ($P_{\beta'}$). At still lower temperatures, other even 
more ordered phases can be identified. Nature usually maintains 
biomembranes in the fluid state. Nevertheless, it is remarkable that 
the transition between the fluid and the gel state, the so-called main 
transition, occurs at temperatures well above room temperature for some 
of the most common lipid molecules, ({\em e.g.}, $\sim 41{}^0$C for the 
phospholipid dipalmitoylphosphatidyncholine (DPPC)) \cites{KC98}.

The gel state of membranes differs from the fluid state in several
respect. As mentioned above, the hydrocarbon chains are much more
ordered {\em i.e.}, they have fewer gauche bonds. As a result, 
the membrane is thicker and the nematic order parameter inside the 
membrane is higher. Furthermore, the molecules also have a 
higher degree of (two-dimensional) positional order within the
membrane: Gel membranes are presumably hexatic or 
(quasi)crystalline \cites{SSS88}.  Both factors, the higher chain order 
and the higher positional order, significantly affect the elastic 
properties of membranes: They are much stiffer in the gel state 
than in the fluid state.

The simplest Ansatz for the curvature elastic free energy
density per unit area of a tensionless fluid membrane is given by
\begin{equation}
\label{eq:helfrich}
{\cal F}/A = 2 \kappa (H -c_0)^2 + \bar{\kappa} K,
\end{equation}
where $2 H$ is the total curvature of the membrane,
{\em i.e.}, the sum of the inverse curvature radii, and $K$ 
is the Gaussian curvature, {\em i.e.}, their product. The
elastic parameters $\kappa$, $\bar{\kappa}$, and $c_0$
are the bending rigidity, the Gaussian rigidity, and the spontaneous
curvature of the membrane\cite{note_c0}. For symmetric bilayers, $c_0$ is zero. 
Furthermore, the Gauss-Bonnet theorem states that the integral over $K$ 
on closed membranes contributes a constant which only depends on the 
topology, hence it can be omitted if the latter is fixed. The
only remaining parameter is the bending rigidity $\kappa$.

In hexatic or crystalline membranes, the situation is more
complex due to the fact that lipids have fixed neighbors
and membranes can sustain elastic shear stress. In a seminal
paper of 1987 \cites{NP87}, Nelson and Peliti have pointed out that 
in this case, fluctuations {\em stiffen} the membranes, and
the effective bending rigidity appears to be increased on 
larger length scales. Out-of-plane thermal undulations with 
different wavevectors ${\bf q}$ are accompanied by in-plane 
shear deformations and thus penalized. This leads to
long-range phonon-mediated interactions between capillary 
waves, which renormalize the free energy (\ref{eq:helfrich}). 
For crystalline membranes, Nelson and Peliti predicted 
that the effective bending rigidity scales as 
$\kappa_R(q) \propto 1/q$ as a function of the 
wavevector. Later, this estimate was refined by 
Le Doussal and Radzihovsky \cites{DR92}, who obtained
$\kappa_R(q) \propto q^{-0.821}$ within a self-consistent
screening approximation. In hexatic membranes, the positional 
order is destroyed due to the presence of free dislocations,
but they still have long range 'bond-orientational' order,
{\em i.e.}, the vectors connecting nearest neighbor
lipids have well-defined average orientations. According 
to Nelson and Peliti, the bending stiffness then still 
increases for large length scales, but only logarithmically,
$\kappa_R(q) \sim \sqrt{- \ln(q \xi_T)}$, where $\xi_T$
is a translational correlation length.  Nelson and Peliti 
also argued that physical lipid membranes should be hexatic 
rather than crystalline, because the ability to buckle 
reduces the elastic energy of free dislocations.

The situation is further complicated in the $L_{\beta'}$ 
phase due to the collective tilt of the molecules. 
Park \cites{P96} has studied fluctuating tilted hexatic 
membranes within a model where two order parameter 
fields, one describing tilt order and one bond-orientational 
order, evolve on a fluctuating fluid surface ({\ref{eq:helfrich}). 
Local shear modes, which play a major role in the bending 
renormalization mechanism of Nelson and Peliti, are not 
included. In this theory, the bending rigidity in the
so-called 'strong coupling limit' renormalizes to a fixed
$q$-independent value, both for untilted 
\cites{PL95} and tilted \cites{P96} membranes.

It should be noted that fluctuations also renormalize the 
bending rigidity in fluid membranes, albeit in the opposite way
than in hexatic/crystalline membranes. On large length scales
the membranes become softer, and $\kappa_R$ effectively
vanishes in the infinite wavelength limit\cites{PL85}. 
However, this softening becomes only relevant on length scales 
where the membranes exhibit large deviations from planar. In practical 
simulation studies of fluctuating self-assembled planar fluid 
membranes, a well-defined bending rigidity that can be associated 
with the unrenormalized (local) bending rigidity can usually be 
extracted from the capillary fluctuation spectrum 
\cites{GGL99,LE00,MM01,LMK03,BB06,WBS09}. 

The purpose of the present work was to study the elastic 
properties of membranes in the fluid and the gel state,
and in particular, to look for signatures of the liquid/gel 
transition in the membrane fluctuation spectrum. To this
end we use a coarse-grained approach to membrane modeling 
\cites{VSK06,S09}. We have carried out extensive simulations of
a very simple generic lipid model \cites{LS05,SDL07}, 
which has been shown to self-assemble into bilayers
with reasonably realistic elastic parameters in the fluid
phase \cites{WBS09} and  to reproduce the most important
membrane phase transitions of DPPC \cites{LS07}, {\em i.e.}, 
the transition between the $L_\alpha$-phase, the $P_{\beta'}$ 
phase, and the $L_{\beta'}$-phase. In previous work, we have 
characterized the fluid phase and analyzed the resulting 
membrane-protein and membrane-mediated protein-protein 
interactions \cites{WBS09}. Now, we will consider the 
$L_{\beta'}$ phase, focussing on the stress profiles in 
the membranes, which can be related to the elastic parameters, 
and on the capillary fluctuations.

The paper is organized as follows: In the next section, we
introduce the model and the simulation method. The results
are presented in section \ref{sec:results}. Finally we
summarize and conclude in section \ref{sec:summary}

\section{Model and Methods}

\label{sec:model}

In our model, each lipid is represented by a chain of seven beads 
with one head bead of diameter $\sigma_h$ followed by six tail beads 
of diameter $\sigma_t$. Non-bonded beads interact, {\em via} a 
truncated and lifted Lennard-Jones potential:
\begin{equation}
  \label{eq:lj_bead}
  V_{\mbox{\tiny bead}}(r) = \left \{
  \begin{array}{rl}
    V_{\mbox{\tiny LJ}}(r/\sigma) - V_{\mbox{\tiny LJ}}(r_c/\sigma) & \mbox{  if $r < r_c$} \\
    0 & \mbox{  otherwise}
\end{array}
\right.
\end{equation}
with
\begin{equation}
  \label{eq:lj}
  V_{\mbox{\tiny LJ}}(x) = \epsilon \left( x^{-12} - 2 x^{-6} \right)
\end{equation}
where $\sigma$ is the mean diameter of the two interacting beads,
$\sigma_{ij} = (\sigma_i + \sigma_j)/2$ ($i,j = h$ or $t$).
Head-head and head-tail interactions are purely repulsive ($r_c = \sigma$),
while tail-tail interactions also have an attractive
contribution ($r_c = 2\sigma$). Bonded beads are connected 
by FENE (Finitely Extensible Nonlinear Elastic) springs with 
the spring potential
\begin{equation}
  V_{\mbox{\tiny FENE}} (r) = -\frac{1}{2} \epsilon_{_{\mbox{\tiny FENE}}} 
     (\Delta r_{\mbox{\tiny max}})^2
  \log \left (1 - \left ( \frac{r - r_0}{\Delta r_{\mbox{\tiny max}}} \right )^2
  \right ),
  \label{eq:fene}
\end{equation}
where $r_0$ is the equilibrium distance, $\Delta r_{\mbox{\tiny max}}$ the maximal
deviation, and $\epsilon_{\mbox{\tiny FENE}}$ the FENE spring constant.
In addition, we introduce a bond-angle potential
\begin{equation}
  \label{eq:ba}
  V_{BA}(\theta) = \epsilon_{BA} (1 - \cos(\theta)).
\end{equation}
The aqueous environment of the membrane is modeled with ``phantom''
solvent beads~\cites{LS05}, which interact with lipids like head
beads ($\sigma_s = \sigma_h$), but have no interactions with each other.

The model parameters are~\cites{DS01,SDL07} $\sigma_h = 1.1 \sigma_t$, 
$r_0 = 0.7\sigma_t$, $\Delta r_{\mbox{\tiny max}} = 0.2 \sigma_t$,
$\epsilon_{_{\mbox{\tiny FENE}}} = 100 \epsilon/\sigma_t^2$, and
$\epsilon_{BA} = 4.7 \epsilon$, and the pressure was chosen
$P = 2. \epsilon/\sigma_t^3$. At these parameters, the lipids 
spontaneously self-assemble into stable bilayers \cites{LS05},
and the bilayer undergoes a main transition to a tilted gel phase 
$L_{\beta'}$ {\em via} a ripple phase $P_{\beta'}$ at the temperature 
$k_B T = 1.2 \epsilon$~\cites{LS07}.

The system was studied using Monte Carlo simulations at 
constant pressure and temperature with periodic boundary conditions 
in a simulation box of variable size and shape: The simulation box 
is a parallelepiped spanned by the vectors $(L_x,0,0), (s_{yx} L_x,L_y,0),
(s_{zx}L_x,s_{zy} L_y,L_z)$, and all $L_i$ and $s_j$ are allowed to
fluctuate. This ensures that the membranes have no interfacial 
tension. We have verified that the total pressure tensor had the
form ${\cal P}_{\alpha \beta} = P \delta_{\alpha \beta}$ with
the applied pressure $P$. We have studied systems of up to $7200$ lipids,
and the run lengths were up to to $8$ million Monte Carlo steps, 
where one Monte Carlo step corresponds to one Monte Carlo move 
per bead, and moves that change the shape of the 
simulation box were attempted every 50th Monte Carlo step. 
The code was parallelized using a domain decomposition scheme 
described in Ref.~\citeonline{SDL07}. The configurations were initially
set up as an ordered bilayers with straight chains in the 
$(xy)$-plane. Typical equilibration times were $1$ million Monte Carlo 
steps for the small systems, which we used to determine pressure 
profiles, and  $4$ million Monte Carlo steps for the large systems, 
which we used to extract fluctuation spectra. After this time, 
the results for the fluctuation spectra did not change any more.
In the fluid state, one can potentially reduce the equilibration and the 
correlation times by applying additional cluster moves that specifically
speed up the long-wavelength modes \cites{F08}. This was not done here,
since these modes were not the most critical ones in the gel state (see 
below), and the length of the runs was mainly dictated by the need to 
accumulate good statistics over the whole range of wavevectors.
It should be noted that in the gel state, lipids never flipflopped 
from one monolayer to the other, hence the number of lipids on 
both sides remained equal throughout the simulation.

The model units can approximately be translated into standard 
SI units by comparing the bilayer thickness ($\sim 6 \sigma_t$ 
in the $L_{\alpha}$ phase and $\sim 7.7 \sigma_t$ in the 
$L_{\beta'}$ phase) and the area per lipid ($\sim 1.4 \sigma_t^2$ 
in the $L_{\alpha}$ phase and $\sim 1 \sigma_t^2$ in the $L_{\beta'}$ 
phase) with the corresponding numbers for real lipid bilayers. 
The numbers in our model roughly reproduce those of DPPC bilayers
if we identify $\sigma_t \sim 6$\AA ~\cites{o_thesis}. 
The energy scale can be estimated by matching the temperatures of the
main transition: $\epsilon \sim 0.36 \cdot 10^{-20}$J.

\section{Results}

\label{sec:results}

\begin{figure}
  \centering
  \includegraphics[width=0.4\textwidth]{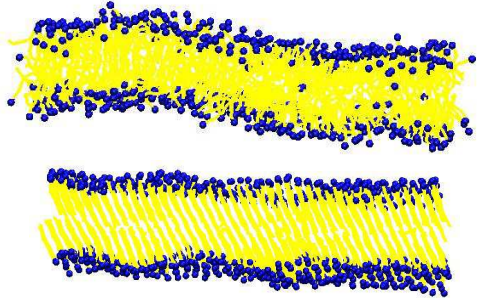} \\
  \caption{Slices through configurations (side view) of a 
  self-assembled lipid bilayer in the fluid phase (top) 
  and the gel phase (bottom).
  }
  \label{fig:snapshots}
\end{figure}

The simulations were mainly carried out at $k_B T = 1.0 \epsilon$, 
which is well in the $L_{\beta'}$ phase. For comparison, we also show 
data for fluid membranes at $k_B T = 1.3 \epsilon$. 
Fig.~\ref{fig:snapshots} shows slices through membrane configurations
in the fluid and the gel state. The corresponding structure factors 
(data not shown) feature distinct peaks in the gel state, corresponding
to triangular positional order, and diffuse Debye-Scherrer rings in the 
fluid state, reflecting lack of positional order in that 
phase \cites{o_thesis}. At the available system sizes, it was not 
possible to determine whether the positional order in the gel phase 
is hexatic or crystalline.

\begin{figure}
  \centering
  \includegraphics[width=0.48\textwidth]{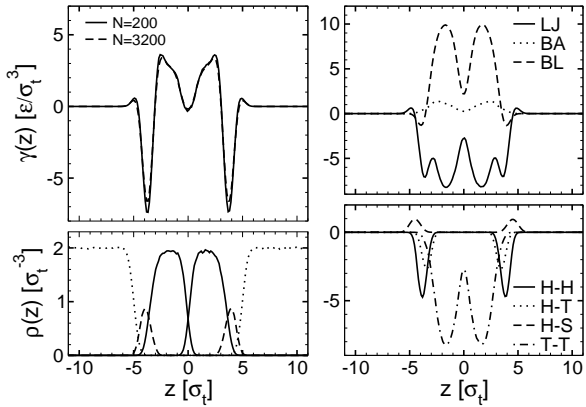} \\
  \caption{Top left: Interfacial tension profile $\gamma(z)$ for lipid bilayers
  with $N=200$ (solid line) and $N=3200$ (dashed line) in the gel phase. 
  In large systems, the profiles are slightly broadened due to the 
  fluctuations of the membrane position. The subsequent profiles 
  were evaluated for systems with $N=200$ lipids.
  Bottom left: Corresponding density profiles of solvent (dotted line), 
  head (dashed lines), and tail beads (solid lines).
  Right: Contributions to the surface tension profile.
  Top right: Contributions from nonbonded interactions 
  (NB, thick solid line), bond angle potentials (BA, dotted line), 
  and bond length potentials (BL, dashed line).
  Bottom right: Among the nonbonded interactions, contributions from
  head-head interactions (HH, solid line), head-tail interaction
  (HT, dotted line), head-solvent interaction (HS, dashed line),
  and tail-tail interaction (TT, dot-dashed line). The contribution 
  from tail-solvent interactions is negligible.
  }
  \label{fig:pressure}
\end{figure}

We begin with considering the local pressure profiles, which give 
information on the internal stress distribution in the membranes. 
The pressure tensor was determined using a version of the 
virial theorem that contains only relative coordinates,
\begin{eqnarray}
\label{eq:pressure}
  \mathcal{P}_{\alpha\beta} &= &
  \frac{N k_B T}{V} \delta_{\alpha\beta} +
\frac{1}{V} \left \langle \sum_{i<j} 
r^\alpha_{ij} F^\beta_{ij;\mbox{\tiny pair}}\right \rangle \\
\lefteqn{- \: 
\frac{1}{V} \left \langle \sum_{j} 
\frac{\mbox{d}V_{BA}(\theta_j)}{\mbox{d}\theta_j}
\Big( r^\alpha_{j-1,j} \frac{\partial \theta_j}{\partial r^\beta_{j-1}} + 
  r^\alpha_{j+1,j} \frac{\partial \theta_j}{\partial r^\beta_{j+1}}
\Big)
\right \rangle,}\qquad \qquad 
\nonumber
\end{eqnarray}
where $N$ is the number of particles, $T$ the temperature, 
$V$ the volume, $\mathbf{r}_{ij} = \mathbf{r}_i - \mathbf{r}_j$ 
the vector connecting the particles $i$ and $j$, 
$\mathbf{F}_{ij:\mbox{\tiny pair}}$ the pairwise force acting on 
particle $i$ from particles $j$, and $\theta_j$ the angle between
the two bonds from bead $j$. To determine the 
local pressure profiles, the system was divided in bins of 
length $\delta z = 0.125 \sigma_t$ along the $z$-axis.
The contributions of relative vectors $\mathbf{r}_{ij}$ 
were parceled out on the bins according to the Irving-Kirkwood 
convention \cites{IK50}. 

Fig.~\ref{fig:pressure} shows profiles of the local interfacial tension 
$\gamma(z)$, defined as the difference $P_N(z) - P_T(z)$ of the 
normal and the tangential pressure profile,
\begin{equation}
\label{eq:tension}
\gamma(z) = \mathcal{P}_{zz}(z) -
(\mathcal{P}_{xx}(z) + \mathcal{P}_{yy}(z))/2,
\end{equation}
for membranes in the gel state. We note that for our
tensionless membranes, the integral $\int \mbox{d}z \: \gamma(z)$ 
vanishes.  The contributions of different forces to the full
profile are shown in Fig.~\ref{fig:pressure} (right).

The main features of the profile can be summarized as follows:
It features (i) maxima in the regions where head 
and tail beads meet, reflecting the interfacial energy of the 
hydrophilic/hydrophobic interface, (ii) minima in the head regions, 
reflecting the tendency of the heads to spread out, and (iii) 
a minimum at the center of the bilayer, reflecting the 'negative 
interfacial tension' or adhesive coupling between the monolayers
({\em i.e.}, a tendency of the monolayer-monolayer interface to 
spread out). These peaks were also found in the stress profiles
of fluid membranes in our \cites{WBS09} and similar \cites{GL98}
coarse-grained models, as well as in atomistic and more reastically 
coarse-grained simulations of fluid lipid bilayers 
\cites{MRY07,note_marrink}. The additional weak maxima at the 
head/solvent interfaces reflect a head-solvent interfacial energy 
which is specific for our model and not always present. 
Overall, the stress profiles in the fluid and the gel
phase are strikingly similar. They differ mainly in amplitude. 
Compared to the fluid state, the surface tension profile 
is amplified by roughly a factor of ten in the gel state, thus
the membrane is under much higher internal tension. Furthermore,
the relative depth of the minima in the head region increases,
compared to the central minima, {\em i.e.}, the spreading pressure
in the head region gains importance. As a result, the
monolayers develop a tendency to bend inwards, {\em i.e.} 
they acquire a spontaneous curvature. 

For fluid membranes, the latter can be estimated by the first 
moment of the stress profile \cite{Safran94},
$\kappa c_0 = - \bar{z}_\gamma$ with
\begin{equation}
\label{eq:kappa_c0}
\bar{z}_\gamma :=  \int_0^{\infty} \mbox{d}z \: 
\gamma_{\mbox{\tiny int}}(z) \: z, 
\end{equation}
where we have used $\int \mbox{d}z \: \gamma(z) = 0$,
and $\kappa/2$ is the monolayer bending rigidity.
We emphasize that $c_0$ refers to the spontaneous curvature 
of a {\em monolayer}. The spontaneous curvature of the 
{\em bilayer} is zero, as mentioned in the introduction. 
In the gel state, the interpretation of $\bar{z}_\gamma$
is less obvious, since the Helfrich description 
(\ref{eq:helfrich}) is no longer valid. As we shall show below,
the very notion of a bending rigidity becomes questionable
on nanometer length scales. Nevertheless, one can argue that
$\bar{z}_\gamma$ still gives qualitative information on
the propensity of monolayers to bend in one direction.

Applying Eq.~ (\ref{eq:kappa_c0}) yields
$\bar{z}_\gamma = - 11.1 \pm 0.1 \epsilon/\sigma_t$. 
In the fluid state, we had obtained
$\bar{z}_\gamma = + 0.3 \pm 0.1 \epsilon/\sigma_t$ in 
Ref. \cite{WBS09}: Thus fluid monolayers 
were found to have a weak tendency to bend outwards, in agreement 
with the results from atomistic simulations \cites{MRY07}.
Here, in the gel state, they have a tendency to bend inwards. 
We conclude that the elastic properties of monolayers in the gel phase 
differ qualitatively from those in the fluid phase.
The interplay of two spontaneous curvatures is presumably one 
of the factors that stabilize a modulated ripple phase in 
the transition region between the fluid and gel phase.

Next we discuss the fluctuation spectra of the membranes.
We have simulated lipid bilayers containing up to 7200 lipids, 
corresponding to patches with areas of roughly 1300 nm${}^2$. 
The data were analyzed following a procedure described in 
Refs.~\citeonline{LMK03,WBS09}. Basically, we determine the 
mean head positions $z_1(x,y)$, $z_2(x,y)$ of the two monolayers 
on a grid $(x,y)$, and derive the height and thickness profiles 
$h(x,y) = (z_1 + z_2)/2$ and $t(x,y) = (z_1 - z_2)/2$.
The spectra are Fourier transformed according to 
\begin{equation}
f_{q_x,q_y} = \frac{L_x L_y}{N_x N_y} \sum_{x,y} f(x,y) 
e^{-i (q_x x + q_y y)},
\end{equation}
and averages $\langle |h_q |^2 \rangle$ and 
$\langle |t_q |^2 \rangle$ are evaluated in $q$-bins 
of size $0.1/\sigma_t$. 

Fig.~\ref{fig:fl} shows the radially averaged results. For comparison, the
corresponding data for the fluid phase (from Ref.~\citeonline{WBS09}) are shown
in the inset. In the fluid phase, the data could be fitted very nicely with an
elastic theory due to Brannigan and Brown\cites{BB06}, which describes the
bilayer in terms of two coupled elastic sheets (monolayers) with two
types of independent fluctuations: ''bending'' fluctuations and ''protrusion''
fluctuations. The spectra are dominated by the protrusion modes at high 
$q$-vectors \cites{GGL99}, and by the bending modes at low $q$-vectors. 

\begin{figure}
  \centering
  \includegraphics[width=0.4\textwidth]{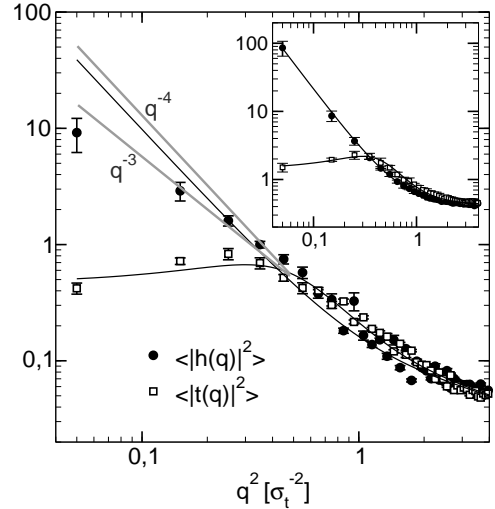}\\
  \caption{Radially averaged Fourier spectrum of the height (full circles)
  and thickness (open squares) fluctuations of membranes in
  the gel state. Black solid lines show the best fit to the elastic theory
  by Brannigan and Brown \cites{BB06} (see text for explanation). 
  Grey solid line indicates the slopes corresponding to a 
  $q^{-4}$-behavior and a $q^{-3}$-behavior. Inset (from 
  Ref.~\protect\citeonline{WBS09}) shows the corresponding data for the 
  fluid membrane for comparison.
  }
  \label{fig:fl}
\end{figure}

In the gel state, the qualitative features of the radially averaged spectra are
at first sight very similar. For example, the height and the thickness spectra
nearly fall on top of each other at high $q$, and the thickness spectrum seems 
to feature a a soft peristaltic mode at $q \sim 0.5 /\sigma_t$, corresponding to a
wavelength around $10 \sigma_t$. Quantitatively, however, one notices distinct
differences between the fluid and the gel state: First, the amplitudes of the
fluctuations in the gel state are much smaller than in the fluid state. Second,
the elastic theory fails at low $q$-vectors, even though it seems to describe
the data reasonably well at high $q$-vectors.  Compared to the theoretical
prediction, the long-wavelength height fluctuations are suppressed. The low-$q$
limit $\langle |h(q)|^2 \rangle \sim q^{-4}$ is never encountered.  The data are
more consistent with the asymptotic behavior predicted by Nelson and Peliti for
crystalline membranes \cites{NP87}, $\langle |h(q)|^2 \rangle \sim 1/\kappa_R(q)
q^{-4} \sim q^{-3}$, but in a double-logarithmic plot, even the slope of
$q^{-3}$ is slightly too steep (Fig.~\ref{fig:fl}).

\begin{figure}[t]
  \centering
  \includegraphics[width=0.4\textwidth]{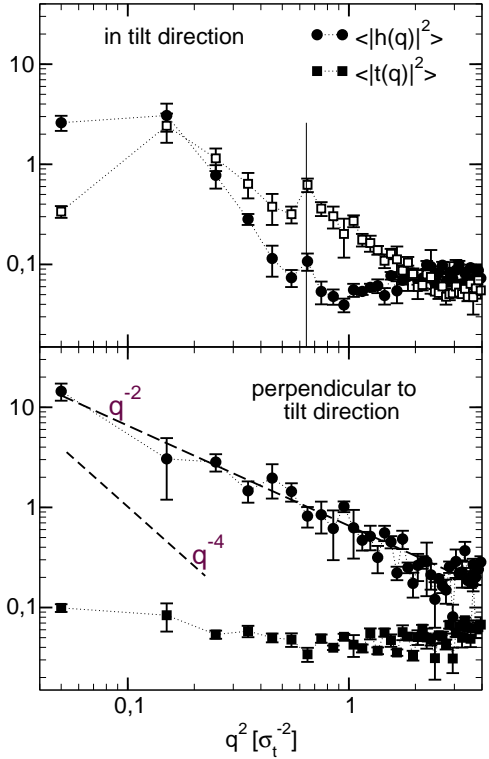}\\
  \caption{
  Fourier spectrum of the height (full circles)
  and thickness (open squares) fluctuations of membranes in
  the gel state for wavevectors parallel (top) and perpendicular
  (bottom) to the tilt direction. 
  Thin vertical bar in the upper panel marks the position
  of a peak in both height and thickness fluctuations.
  Dashed lines in the lower panel show $q^{-2}$-slope
  and $q^{-4}$-slope as indicated. 
  }
  \label{fig:fl_xy}
\end{figure}

To explore this phenomenon in more detail, we have analyzed separately the
fluctuations in tilt direction and perpendicular to the tilt direction. Since
the tilt direction slowly fluctuated throughout the simulations, this means that
before averaging $\langle |h(\mathbf{q})|^2 \rangle$ and $\langle
|t(\mathbf{q})|^2 \rangle$, every configuration had to be rotated such that the
tilt points along a given axis.  The results are shown in Fig. \ref{fig:fl_xy}.
Both in the direction parallel and perpendicular to the tilt, the thickness and
height fluctuation spectra differ strongly from the radially averaged spectra,
and are no longer compatible with the elastic theory.

In the tilt direction, the thickness fluctuation amplitudes are largely
comparable to those of the radially averaged spectrum, with the exception
of the $q$-region around $(q\sigma_t)^2 \sim 0.15$, where they exhibit
a maximum. In contrast, the height fluctuations are strongly suppressed, 
most notably for large $q$-vectors beyond $(q \sigma_t)^2 \sim 0.7$, but also 
for small $q$-vectors. Both the thickness and the height fluctuation spectra 
feature small but distinct peaks at $q \sigma_t \sim 0.8$,
which clearly rise above the range of statistical 
fluctuations. These peaks correspond to undulations with wavelength 
$\lambda \sim 8 \sigma_t$. The tendency of the gel membrane to exhibit 
structure on that length scale can already be sensed when looking at
the bare snapshots of the gel phase (Fig.~\ref{fig:snapshots}).  
The other maximum at $(q \sigma_t)^2 \sim 0.15$ (or
$q \sigma_t) \sim 0.4$) in the thickness spectrum,
which was mentioned in the beginning, corresponds to undulations
with twice the wavelength, $\lambda \sim 16 \sigma_t$. 
In the height spectrum, one can vaguely discern a 'peak' in the same 
$q$-region, which however interferes with the general growth of 
fluctuation amplitudes for small $q$-vectors.

Hence we conclude that both the thickness and the height fluctuation spectra 
feature soft modes with wavelength $\lambda_0 \sim 8 \sigma_t$ 
and $\lambda_1 = 2 \lambda_0$. The wavelength $\lambda_1 = 16 \sigma$
happens to coincide with the period of the ripple phase \cites{LS07}.  
This suggests that the fluctuations in tilt direction already carry 
the signature of the modulated phase which intrudes between the
liquid and the gel state.

In the direction perpendicular to the tilt, the situation
is different. The thickness fluctuations are almost entirely
suppressed. The height fluctuations rise smoothly and
are best compatible with a $q^{-2}$ slope, suggesting that
they are controlled by an interfacial tension rather than a 
bending rigidity. On the other hand, the evaluation of the
pressure tensor clearly shows that the total interface
tension, integrated over the membrane, vanishes within
the error ($\gamma = 0 \pm 0.1 \epsilon/\sigma_t^2$).

An alternative potential explanation for the observed $q^{-2}$ 
shape of the fluctuation spectrum can be derived based on the
following two assumptions: (i) The membrane undulations in the 
direction perpendicular to the tilt (denoted $x$ hereafter)
are dominated by out-of-layer fluctuations, {\em i.e.}, the 
lipids mainly move up and down, keeping their orientation, 
such that energy contributions due to bending or splay are 
negligible (see Fig.~\ref{fig:fluct_modes}). (ii) The effective 
interfacial tension in $x$-direction $\gamma_x$ depends on the 
angle $\theta$ between the projections onto the $xz$ plane
of the lipid orientation vector and the local interface normal.
(We note that this second assumption implies that the $x$-direction 
can be defined independently of the tilt direction, {\em e.g.}, 
in terms of the hexatic order in the membrane.) For small angles
$\theta$, the function $\gamma_x$ can then be expanded in 
powers of $\theta$, $\gamma_x[\theta] \approx A \theta^2/2$.
For constant lipid orientation, the angle $\theta$ depends 
on the local membrane height position $h(x,y)$ {\em via} 
$\theta \approx \partial h/\partial x$. 
Hence the fluctuations of $h(x,y)$ in $x$-direction contribute
to the total free energy with the term
$\int {\rm d}x \: \gamma_x[\theta(x)] 
\approx A/2 \int {\rm d}x \: (\partial h/\partial x)^2$, 
which looks formally like an interfacial tension contribution
with tension $A$. This explains how out-of-layer fluctuations
can display a tension-like behavior. One would expect the 
out-of-layer fluctuations to eventually give way to 
bending-dominated fluctuations at very long wavelengths, 
but the latter are not observed in our simulations.

\begin{figure}
  \centering
  \includegraphics[width=0.45\textwidth]{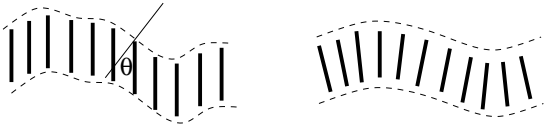}\\
  \caption{
  Schematic sketch of different types of membrane height fluctuations.
  Left: Out-of-layer fluctuations, governed by a surface energy that
  depends on the local (projected) tilt angle $\theta$.
  Right: Bending-dominated height fluctuations, governed by a bending energy.
  }
  \label{fig:fluct_modes}
\end{figure}

\section{Discussion and Summary}

\label{sec:summary}

To summarize, we have analyzed the elastic behavior and the 
fluctuations of lipid membranes in the gel state. 
They are distinctly different from those in the fluid state. 
Whereas the liquid state can be described nicely by an elastic 
theory for fluid coupled monolayers, this description fails in 
the gel state. The failure can be attributed to the ordered
structure of the gel membrane and to its anisotropy.
Our simulation results disclose a complex fluctuation spectrum, 
which cannot be understood within any of the available
continuum theories for membranes.

The spectrum is anisotropic and features two soft modes that 
are very likely related to the ripple phase, which is known to 
emerge in the transition region between the liquid and the gel 
phase. Given the fact that the latter has a periodic modulation
with wavelength $\sim 10$ nm, we must conclude that our systems,
which have linear dimensions of around $\sim 35$ nm, are still 
much too small to give insight into the asymptotic elastic 
behavior of the membranes. Unfortunately, the simulation of much 
larger systems (at least a factor of ten would be necessary)
is unfeasible with our currently available computer resources. 
Our simulations can be used to analyze the elastic behavior of 
the membrane on the length scale of tens of nanometers. Based on 
our data, it is clear that the membrane in the gel state still has 
no single unique 'bending rigidity' on this lateral length scale, 
even though it exceeds the molecular scale (the monolayer
thickness) by more than an order of magnitude. Rather, the 
elastic behavior results from a complex interplay of different 
contributions which cannot easily be told apart. Formulating
an elastic theory for such membranes will be a challenge
for future theoretical work, which should presumably also
contribute to a better understanding of the mechanisms
driving the formation of the ripple phase.

In the present work, we have discussed 'monodomain' gel membranes
with a uniform tilt direction. This seems realistic, given
that scanning optical microscope studies of supported 
DPPC-bilayers in the gel phase \cite{LDW03} have shown that 
single-tilt domains in such membranes extend over 1-2 micrometers. 
On larger scales, the membranes are effectively 'polycrystalline',
and their fluctuations will be dominated by processes at the
domain boundaries and/or defects in the tilt order. Preliminary 
studies, where we have enforced such defects, have shown that 
they amplify the fluctuations considerably. It will be 
interesting to study these effects in more detail in the future. 

\section{Acknowledgments}

We thank J\"org Neder and Olaf Lenz for useful discussions. The simulations
were carried out at the Paderborn center for parallel computing
(PC2) and the John von Neumann Institute for Computing
in J\"ulich. This work was funded by the German Science Foundation
(DFG) within the SFB 613.

\bibliographystyle{rsc}
\bibliography{paper}

\begin{mcitethebibliography}{32}
\bibitem[2()]{note_c0}
We should note that there is a factor-of-two convention disagreement in the
  literature regarding the definition of $c_0$. Here and throughout this paper,
  we use the convention where it is subtracted from the mean curvature.
  (Eqs.~(\protect\ref{eq:helfrich}) and (\protect\ref{eq:kappa_c0})).\relax
\mciteBstWouldAddEndPunctfalse
\mciteSetBstMidEndSepPunct{\mcitedefaultmidpunct}
{}{\mcitedefaultseppunct}\relax
\EndOfBibitem
\end{mcitethebibliography}

\end{document}